\documentclass[prl,12pt]{revtex4}
\usepackage{amssymb}
\usepackage{amsmath}
\usepackage{graphicx}
\usepackage{color}
\DeclareGraphicsExtensions{.jpg,.jpeg,.pdf,.png,.mps,.eps,.ps}

\newcommand{\beq}{\begin{equation}}
\newcommand{\eeq}{\end{equation}}
\newcommand{\bea}{\begin{eqnarray}}
\newcommand{\ena}{\end{eqnarray}}
\newcommand{\etal}{{\it et al.}}
\newcommand{\ie}{{\it i.e.}}
\newcommand{\eg}{{\it e.g.}}
\newcommand{\etc}{{\it etc.}}
\newcommand{\lsim}{\mathrel{\mathop{\kern 0pt \rlap
{\raise.2ex\hbox{$<$}}}
\lower.9ex\hbox{\kern-.190em $\sim$}}}
\newcommand{\gsim}{\mathrel{\mathop{\kern 0pt \rlap
{\raise.2ex\hbox{$>$}}}
\lower.9ex\hbox{\kern-.190em $\sim$}}}

\newcommand{\pslash}[1]{\not\!\!{#1}}

\newcommand{\hepph}[1]{{\tt hep-ph/#1}}

\newcommand{\prep}[3]{Phys.\ Rep.\ {\bf #1}, #3 (#2)}
\newcommand{\plb}[3]{Phys.\ Lett.\ B\ {\bf #1}, #3 (#2)}

\newcommand{\pr}[3]{Phys.\ Rev.\ {\bf #1}, #3 (#2)}
\renewcommand{\prl}[3]{Phys.\ Rev.\ Lett. {\bf #1}, #3 (#2)}
\renewcommand{\prd}[3]{Phys.\ Rev.\ D\ {\bf #1}, #3 (#2)}
\renewcommand{\pra}[3]{Phys.\ Rev.\ A\ {\bf #1}, #3 (#2)}

\newcommand{\href}[2]{#1}

\definecolor{cyan}{cmyk}{1.,0.,0.,0.5}
\definecolor{magenta}{cmyk}{0.,1.,0.,0.5}
\definecolor{verdatre}{cmyk}{0.5,0.,0.5,0.5}
\definecolor{yellow}{cmyk}{0.,0.,0.2,0.0}
\definecolor{rouge}{cmyk}{0.,0.4,0.6,0.0}
\definecolor{orange}{cmyk}{0.,0.5,0.5,0.}
\definecolor{violet}{rgb}{0.5,0.,0.5}

\begin{document}

\noindent
\title{Puzzles of Divergence and Renormalization \\ in Quantum Field Theory}
 \vskip 1.cm
\author{Guang-jiong Ni $^{\rm a,b}$}
\email{\ pdx01018@pdx.edu}
\affiliation{$^{\rm a}$ Department of Physics, Portland State University, Portland, OR97207, U. S. A.\\
$^{\rm b}$ Department of Physics, Fudan University, Shanghai, 200433, China}

\author{Jianjun Xu $^{\rm b}$}
\email{\ xujj@fudan.edu.cn}
\affiliation{$^{\rm b}$ Department of Physics, Fudan University, Shanghai, 200433, China}

\author{Senyue Lou $^{\rm c,d}$}
\email{\ sylou@sjtu.edu.cn}
\affiliation{$^{\rm c}$ Department of Physics, Shanghai Jiao Tong University, Shanghai, 200030, China\\
$^{\rm d}$ Department of Physics, Ningbo University, Ningbo 315211, China}

\vskip 0.5cm
\date{\today}

\vskip 0.5cm
\begin{abstract}
A regularization renormalization method ($RRM$) in quantum field theory ($QFT$) is discussed with simple rules:
Once a divergent integral $I$ is encountered, we first take its derivative with respect to some mass parameter enough
times, rendering it just convergent. Then integrate it back into $I$ with some arbitrary constants appeared. Third, the
renormalization is nothing but a process of reconfirmation to fix relevant parameters (mass, charge, \etc) by experimental data via suitable choices of these constants. Various $QFT$ problems, including the Lamb shift, the running coupling constants in $QED$ and $QCD$, the $\lambda \phi^4$ model as well as Higgs mass in the standard model of
particle physics, are discussed. Hence the calculation, though still approximate and limited in accuracy, can be performed
in an unambiguous way with no explicit divergence, no counter term, no bare parameter and no arbitrarily running mass scale (like the $\mu$ in $QCD$).

\end{abstract}

\maketitle \vskip 1cm

\section{Introduction}
\label{sec:introduction}

\vskip 0.1cm

Since the establishment of quantum mechanics ($QM$) and the quantization of electromagnetic field, the quantum electrodynamics ($QED$) in particular and the quantum field theory ($QFT$) in general have been developed for over 80 years. A common prominent feature of $QFT$ is the emergence of the divergence in calculations beyond the tree level. To handle
these divergences, various regularization and renormalization methods ($RRMs$) have been proposed.

Despite the great success of $QFT$, the present status of $RRMs$ remains ambiguous to some extent. For example, in the
theory of Chromodynamics ($QCD$) for describing the strong interactions of colored quarks and gluons, a commonly used
renormalization scheme ($RS$) is the modified minimal substraction ($\bar{MS}$) scheme (see a summary in the Review of
Particle Physics in 2008, \cite{1}, p.157) where an arbitrary renormalization mass scale $\mu$ is introduced (see next
section). Physicists believe the fundamental theorem of $RS$ dependence: Physical quantities, such as the cross-section
calculated to all orders in perturbation theory, should not depend on the $RS$. However, it follows that a truncated
series does exhibit $RS$ dependence. In practice, $QCD$ cross-section are known to different orders, depending on the
choice of $RS$ (and $\mu$) in different sensitive ways. We still don't know what is the "best" choice for $\mu$ within
a given scheme (usually $\bar{MS}$). There is no definite answer to this question yet.

\section{What a divergence means?}
\label{sec:divergence}

\vskip 0.1cm

Physicists often talk about different orders of a divergence, based on its dimension with respect to mass (\ie, momentum,
we use natural unit system with $\hbar=c=1$). For example, if a Feynman diagram integral ($FDI$) in $QFT$ reads
\begin{equation}\label{1}
I=\int\dfrac{d^4K}{(2\pi)^4}\dfrac{K^n}{(K^2-M^2)^2}
\end{equation}
where $K$ corresponds to the (4-dimensional)momentum of virtual particle (say, virtual photon in $QED$) and $M$ is a
mass parameter (maybe in a complex form) characterizing the $QFT$ under consideration. Then
\begin{equation}\label{2}
n=\left\{\begin{array}{l}
           0,\quad \text{logarithmic divergence} \\
           1,\quad \text{linear divergence} \\
           2,\quad \text{quadratic divergence}
         \end{array}\right.
\end{equation}
However, among these categories, only the first one is really meaningful in mathematics. This is because the definition
of a number sequence $A_i\;(i=1,2,\ldots)$ having a limit being $\infty$ is as follows: Given arbitrarily a large
number $M$, one can always find such a number $N$ so that $A_i>M$, when $i>N$. Here $A_i$ and $M$, let alone $i$ and $N$,
are all dimensionless numbers. A number $M\gg1$ is called a large number, whereas $\varepsilon\ll1$ a small number.

On the other hand, the space-time coordinates $\bf x$ and $t$, mass $m$ and momentum $p$ (or $k$) are physical quantities and each with certain dimension. If treating them as dimensionless numbers, we will run into trouble inevitably.

Example A: Assume that in $QFT$, a $FDI$ with linear divergence is approximately expressed as $I\sim 10^3M$ with $M$ being a
mass parameter in the unit of $mg$. If $M=1\,mg$, we have $I\sim 10^3mg$ with $10^3\gg1$ being a large number. But if changing the unit from $mg$ to $kg$, we will have $I\sim 10^{-3}kg$ with $10^{-3}\ll1$ being a small number. A mathematician
would ask: "Could you still treat your $I$ as a divergent quantity?"Who could answer his question?

Example B: In the $\bar{MS}$ scheme, the renormalization mass scale $\mu$ is introduced as follows (see p.137 in \cite{2})
\begin{equation}\label{3}
\dfrac{\Gamma(2-d/2)}{(4\pi)^{d/2}(m^2)^{2-d/2}}=\dfrac{1}{(4\pi)^2}\left[\dfrac{2}{\varepsilon}-\gamma
+\ln(4\pi)-\ln(m^2)\right]\to \dfrac{1}{(4\pi)^2}\left[-\ln\left(\dfrac{m^2}{\mu^2}\right)\right]
\end{equation}
where $m$ is some mass parameter containing in the model. Eq.(\ref{3}) is derived from the "dimensional regularization"
method. The 4-dimensional (Euclidean) space has been analytically continued into $d$-dimensional one with $\varepsilon\sim 4-d\sim 0$
and Gamma function $\Gamma(2-d/2)=\Gamma(\varepsilon/2)=2/\varepsilon-\gamma$ ($\gamma=0.4772\ldots$ is the Euler constant).
Obviously, the left-handed-side ($LHS$) has a dimension of $m^{-\varepsilon}$, whereas in the right-handed-side ($RHS$),
the function $\ln\left(\frac{m^2}{\mu^2}\right)$ becomes dimensionless after the $\mu$ is introduced. However, the mathematician would feel quite uncomfortable because $\varepsilon\neq0$. He will focus on the middle of Eq.(\ref{3}) and
ask: "Why the divergent number $2/\varepsilon$ disappears at the $RHS$ and becomes finite? Where the term $\ln(m^2)$ comes
from? What is its dimension?"

We physicists accept Eq.(\ref{3}) since it could be derived from a "mathematical formula" like (see p.57 in \cite{3})
\begin{equation}\label{4}
(m^2)^{d/2-2}=\exp\left[\left(\frac{d}{2}-2\right)\ln m^2\right]\simeq 1+\left(\frac{d}{2}-2\right)\ln m^2
\end{equation}
Then the mathematician would say: "No! In the mathematical formula
\begin{equation}\label{5}
x^y=\exp[y\ln x]
\end{equation}
both $x$ and $y$ must be dimensionless numbers. So in the $LHS$ of Eq.(\ref{3}), you correctly write down:
\begin{equation*}\label{}
(4\pi)^{-d/2}=(4\pi)^{2-d/2-2}=\dfrac{1}{(4\pi)^2}\exp\left[(2-\frac{d}{2})\ln(4\pi)\right]\simeq
\dfrac{1}{(4\pi)^2}\left[1+(2-\frac{d}{2})\ln(4\pi)\right]
\end{equation*}
But Eq.(\ref{4}) is wrong because $x=m^2$ is a physical quantity with dimension. That's why you got a strange result at the $RHS$ of Eq.(\ref{3})".

Hence if insisting on mathematical rigor, we should admit that the introduction of $\mu$ via Eq.(\ref{3}) is groundless.
Then a question arises: Why the $\mu$ seems necessary in $QCD$ ?

To our understanding, the answer lies in the fact that in high energy $QCD$, the quarks' masses were often neglected. Therefore, in order to express the running coupling constant ($RCC$) of strong interaction, $\alpha_s$, as a function of
$Q$, the 3-dimensional momentum transfer in collision, one needs $\mu$ as shown by the solution of renormalization-group-equation ($RGE$) (see p.532 of \cite{2} and Eq.(\ref{53}) below):
\begin{equation}\label{6}
\alpha_s(Q)=\dfrac{\alpha_s(\mu)}{1+\alpha_s(\mu)\frac{\beta_0}{2\pi}\ln(Q/\mu)}
\end{equation}
It is interesting to solve Eq.(\ref{6}) for $\alpha_s(\mu)$, yielding
\begin{equation}\label{7}
\alpha_s(\mu)=\dfrac{\alpha_s(Q)}{1+\alpha_s(Q)\frac{\beta_0}{2\pi}\ln(\mu/Q)}
\end{equation}\footnotemark[1]
\footnotetext[1]{see Appendix of \cite{4}, where a typing error exists in the denominator of Eq.(A.5), "$+$" should be "$-$".}
We see that Eqs.(\ref{6}) and (\ref{7}) are symmetrical with respect to mutual change of $Q\leftrightarrow \mu$. $Q$ and $\mu$ are
essentially equivalent. Why we need both of them? The answer is: only $Q/\mu$ is capable of expressing a dimensionless
$\alpha_s$. However, as shown in Eq.(\ref{3}), the existence of $\mu$ is doubtful, even superfluous. Once we take the quarks' masses into account, there will be no need of $\mu$ at all (see section \ref{sec:hadronization} below).

\section{Self-Energy Correction of an Electron, Lamb Shift}
\label{sec:selfenergy}

\vskip 0.1cm

As is well known (see \eg, Refs.\cite{4,5,6}), the $FDI$ of a free electron's self-energy at one loop ($L=1$) level of $QED$
in covariant form reads
\begin{equation}\label{8}
-i\Sigma(p)=(ie)^2\int\dfrac{d^4k}{(2\pi)^4}\dfrac{g_{\mu\nu}}{ik^2}\gamma^\mu\dfrac{i}{\pslash{p}-\pslash{k}-m}\gamma^\nu
\end{equation}
where the Bjorken-Drell metric ($\pslash{p}=\gamma^\mu p_\mu$) and rationalized Gaussian units are adopted with electron charge $-e\,(e>0)$ and mass $m=m_e$. In Eq.(\ref{8}), $p$ and $k$ are momenta of electron and (virtual) photon. After
introducing the Feynman parameter $x$ and making a shift in momentum integration: $k\to K=k-xp$, Eq.(\ref{8}) is recast into
\begin{equation}\label{9}
-i\Sigma(p)=-e^2\int_0^1dx[-2(1-x)\pslash{p}+4m]I
\end{equation}
with
\begin{equation}\label{10}
I=\int\dfrac{d^4K}{(2\pi)^4}\dfrac{1}{(K^2-M^2)^2},\quad M^2=p^2x^2+(m^2-p^2)x
\end{equation}
being a logarithmically divergent integral, see Eq.(\ref{1}).

Note that in Eq.(\ref{10}) we can change the unit of $M$ (and $K$) at our disposal without any change in the value of $I$,
which is just a "dimensionless", "large" but "uncertain" number. However, in the past, we used to pay too much attention to its feature of being "large", trying to curb the divergence by means of some regularization method, which led to complicated
renormalization schemes ($RS$).

By contrast, now we believe the more important, even essential feature of a divergence is hiding in its "uncertainty".
To stress this cognition, we just use a simple trick to regulate the $I$ in Eq.(\ref{10}) as follows.

To render it convergent, we perform a differentiation with respect to the mass-square parameter $M^2$, yielding
\begin{equation}\label{11}
\dfrac{\partial I}{\partial M^2}=2\int\dfrac{d^4K}{(2\pi)^4}\dfrac{1}{(K^2-M^2)^3}=\dfrac{-i}{(4\pi)^2}\dfrac{1}{M^2}
\end{equation}
Then we reintegrate Eq.(\ref{11}) with respect to $M^2$ and arrive at
\begin{equation}\label{}
I=\dfrac{-i}{(4\pi)^2}(\ln M^2+C_1)=\dfrac{-i}{(4\pi)^2}\ln\dfrac{M^2}{\mu_2^2}
\end{equation}
where an arbitrary constant $C_1=-\ln \mu_2^2$ (with $\mu_2$ a mass scale to be fixed later) is introduced so that the
ambiguity of dimension in the $\ln M^2$ term can be eliminated.

Further integration of Eq.(\ref{9}) with respect to $x$ leads to ($\alpha=\frac{e^2}{4\pi}$)
\begin{equation}\label{13}\begin{array}{l}
\Sigma(p)=A+B\pslash{p}, \\[5mm]
A=\dfrac{\alpha}{\pi}m\left[2-2\ln\dfrac{m}{\mu_2}+\dfrac{(m^2-p^2)}{p^2}\ln\dfrac{(m^2-p^2)}{m^2}\right], \\[5mm]
B=\dfrac{\alpha}{4\pi}\left\{2\ln\dfrac{m}{\mu_2}-3-\dfrac{(m^2-p^2)}{p^2}\left[1+\dfrac{(m^2+p^2)}{p^2}
\ln\dfrac{(m^2-p^2)}{m^2}\right]\right\}
\end{array}
\end{equation}
Using the chain approximation, we can derive the modification on the electron propagator as
\begin{equation}\label{14}
\dfrac{i}{\pslash{p}-m}\to \dfrac{i}{\pslash{p}-m}\dfrac{1}{1-\frac{\Sigma(p)}{\pslash{p}-m}}=\dfrac{iZ_2}{\pslash{p}-m_R}
\end{equation}
where
\begin{equation}\label{15}
Z_2=\dfrac{1}{1-B}
\end{equation}
is the renormalization factor for electron's wave function and
\begin{equation}\label{16}
m_R=\dfrac{m+A}{1-B}
\end{equation}
is the renormalized mass of $m$. The increment of mass reads
\begin{equation}\label{17}
\delta m=m_R-m=\dfrac{A+mB}{1-B}
\end{equation}
In the past, many physicists viewed $\delta m$ as some real contribution of "radiation correction". While $m_R$ should be
identified with the observed mass $m_{obs}$, or physical mass $m_e$, the original $m$ (usually denoted by $m_0$ or $m_B$ in
the expression of Lagrangian density) was thought to be a "bare mass". Both $\delta m$ and $m_0$ were divergent quantities.
(see, \eg, p.220 in \cite{2}).

We don't think so. Let us read carefully the seminal paper by Bethe in 1947\cite{8}. The theory for the hydrogenlike atom begins with a Hamiltonian in the center-of-mass frame
\begin{equation}\label{18}
H_0=\dfrac{{\bf p}^2}{2m}+\dfrac{{\bf p}^2}{2m_N}-\dfrac{Z\alpha}{r}=\dfrac{{\bf p}^2}{2\mu}-\dfrac{Z\alpha}{r}
\end{equation}
Bethe pointed out that the effect of electron's interaction with the vector potential $\bf A$ of radiation field
\begin{equation}\label{19}
H_{int}=\dfrac{e}{mc}{\bf A}\cdot{\bf p}
\end{equation}
should properly be regarded as already included in the $m_{obs}$, which is denoted by $m$ in Eqs.(\ref{18}) and (\ref{19}).

In our understanding on Bethe's claim, the "self-interaction" of electron with radiation field is indivisible from the free
electron mass $m$. In other words, in the covariant form of $QED$, certain contributions of $FDIs$ for "self-energy" (with
Eq.(\ref{8}) being merely that at $L=1$ order) at all orders (up to $L\to \infty$) are already contained in the value of $m$. To show this cognition, we impose the mass-shell condition $p^2=m^2$ in Eq.(\ref{17}) together with
\begin{equation}\label{20}
\delta m |_{p^2=m^2}=\dfrac{\alpha m}{4\pi}(5-6\ln\frac{m}{\mu_2})=0
\end{equation}
which in turn fixes the arbitrary constant $\mu_2$ to be
\begin{equation}\label{21}
\mu_2=me^{-5/6}
\end{equation}
and thus
\begin{equation}\label{22}
Z_2|_{p^2=m^2}=\dfrac{1}{1+\frac{\alpha}{3\pi}}\simeq 1-\dfrac{\alpha}{3\pi}
\end{equation}
Note that $m_R=m=m_{obs}=m_e$ with no bare mass at all and $Z_2$ is fixed and finite, in sharp contrast to that in
previous theories.

Our reader may wonder: "In this case, does the calculation on $FDI$ for the self-energy become worthless ?" The answer
is "No" due to two reasons. First, at the $QM$ level, the parameters $m$ and $e$ in Eqs.(\ref{18}) and (\ref{19}) can be
regarded as well-defined. But they are not so at the level of $QED$. As discussed before Eq.(\ref{20}), the new effect of
radiative corrections of $FDIs$ for self-energy is inevitably confused with that in the mass, the dividing line between
them is blurred. In some sense, the appearance of divergence in the $FDI$ is just a warning: the new effect you want to calculate has become entangled with the mass $m$, rendering both of them uncertain. Hence the aim of so-called mass
renormalization is nothing but a reconfirmation of $m$ as we did in Eqs.(\ref{20})-({22}), where the mass $m$ is renormalized on the mass-shell $p^2=m^2$ with $m=m_e$ being fixed by the experimental value and thus well-defined. This is one important thing we must do and at most we can do on the mass-shell for a free electron.

Second, the increment of mass, $\delta m$, ceases to be zero once when the electron is moving off-mass-shell ($p^2\neq m^2$). Then Eq.(\ref{17}) will provide some information about the new effect of radiation corrections. For example, for a bound electron in a hydrogenlike atom, in Ref.\cite{7}, we replace the electron mass $m=m_e$ by reduced mass $\mu=\frac{m_em_N}{m_e+m_N}$ (not to be confused with the $\mu$ in $QCD$) and write (see also \cite{29}):
\begin{equation}\label{23}
p^2=\mu^2(1-\zeta)
\end{equation}
Here a dimensionless parameter $\zeta\,(>0)$ is introduced to show (on average) how large the extent of "off-mass-shell" is.
Substitution of Eq.(\ref{23}) into Eq.(\ref{17}) yielding
\begin{equation}\label{24}
\delta \mu\simeq \dfrac{\alpha \mu}{4\pi}\dfrac{(-\zeta+2\zeta\ln\zeta)}{1+\frac{\alpha}{3\pi}}
\end{equation}
where some terms of the order of $\zeta^2$ or $\zeta^2\ln\zeta$ are neglected since $\zeta\ll1$.

As a perturbative calculation at $L=1$ order, we may ascribe $\delta\mu$ to the (minus) binding energy $B$ of electron
in the Bohr theory
\begin{equation}\label{25}
\delta \mu=\varepsilon_n=-B=-\dfrac{Z^2\alpha^2}{2n^2}\mu
\end{equation}
Combination of Eqs.(\ref{24}) and (\ref{25}) gives the value of $\zeta=\zeta^{<S>}$ with the superscript $<S>$ referring to
"self-energy (at $L=1$ order)".

Another "nonperturbative" method to fix the $\zeta$ in Eq.(\ref{23}) is to resort to the Virial theorem: For an electron in the Coulomb potential $V=-\frac{Z\alpha}{r}$, its kinetic energy $T=\frac{1}{2\mu}{\bf p}^2$ can be evaluated on average as
\begin{equation}\label{26}
<{\bf p}^2>=2\mu<T>=2\mu[-B-<V>]=2\mu B
\end{equation}
\begin{equation}\label{27}
<p^2>=<E^2-{\bf p}^2>=<(\mu-B)^2-{\bf p}^2>\simeq \mu^2(1-\dfrac{4B}{\mu})
\end{equation}
Comparing Eq.(\ref{27}) with Eq.(\ref{23}), we obtain
\begin{equation}\label{28}
\zeta^{<V>}=\dfrac{4B}{\mu}=\dfrac{2Z^2\alpha^2}{n^2}
\end{equation}
where the superscript $<V>$ refers to "Virial theorem".

In Ref.\cite{7}, for explaining the Lamb shift of energy levels in hydrogenlike atoms, we find the result being expressed
in terms of $\zeta$. Throughout the entire calculation, all ultraviolet divergences are handled like that in Eqs.(\ref{9})-({12}) while the infrared divergence disappears due to the introduction of $\zeta$. However, the formulas
are still approximate and either one of $\zeta^{<S>}$ and $\zeta^{<V>}$ is not reliable. So in the following table $I$,
not only $\zeta^{<S>}$ and $\zeta^{<V>}$, but also two kinds of "average", $\zeta^{<S+V>}=\frac{1}{2}(\zeta^{<S>}+\zeta^{<V>})$ and $\zeta^{<SV>}=\sqrt{(\zeta^{<S>}\zeta^{<V>}}$ are given.

\vskip 0.5cm
\begin{small}\hspace*{-16mm}\begin{tabular}{|c|c|c|c|c|c|c|c|c|}
\multicolumn{8}{c}{Table I. Off-mass-shell parameter $\zeta$ and $\ln\zeta$}\\[5pt]
  \hline
  $\frac{Z^2}{n^2}$ & $\zeta^{<S>}\times 10^4$ &-$\ln\zeta^{<S>}$ & $\zeta^{<V>}\times 10^6$ & -$\ln\zeta^{<V>}$ &
  $\zeta^{<S+V>}\times 10^5$ & -$\ln\zeta^{<S+V>}$ & $\zeta^{<SV>}\times 10^5$ & $-\ln\zeta^{<SV>}$ \\
  \hline
  $\frac{1}{16}$ & $1.546093458$ & $8.77461$ & $\frac{\alpha^2}{8}=6.6564192$
   &11.91992886 & $8.0632$ & 9.425609 & 3.2080284 & 10.34727 \\
   \hline
  $\frac{1}{4}$ & $7.446539697$ & 7.20259 & $\frac{\alpha^2}{2}=26.6256771$
  & 10.5336345 & $38.5639$ & 7.860609 & 14.0808 & 8.86816225 \\
   \hline
 1 & $37.73719345$ & 5.57969 & $2\alpha^2=106.502$ &9.147340142 &
 $194.011$ & 6.2450103 & 63.39626 & 7.36351521 \\
  \hline
\end{tabular}\end{small}
\vspace{0.5cm}
\normalsize

There are $8$ cases discussed in \cite{7}. The first one is the hydrogen atom's "classical Lamb shift" measured as:
\begin{equation}\label{29}
L_H^{exp}(2S-2P)=E_H(2S_{1/2})-E_H(2P_{1/2})=1057.845\;MHz
\end{equation}
Theoretically, the radiative correction (at $L=1$ order) makes the dominant contribution, yielding:
\begin{equation}\label{30}\begin{array}{l}
\Delta E_H^{Rad<S>}(2S-2P)=1000.657\;MHz\\
\Delta E_H^{Rad<S+V>}(2S-2P)=1089.651\;MHz\\
\Delta E_H^{Rad<SV>}(2S-2P)=1226.087\;MHz\\
\Delta E_H^{Rad<V>}(2S-2P)=1451.791\;MHz
\end{array}
\end{equation}
Taking the small contribution from the nuclear size effect into account, we adopt the $<S+V>$ scheme to obtain
\begin{equation}\label{31}
L_H^{theor}(2S-2P)=1089.794\;MHz
\end{equation}
which is larger than the experimental value, Eq.(\ref{29}), by $3\%$.

The most interesting case is the $1S-2S$ two-photon transition in hydrogen $H$ or deuterium $D$ because its natural width
is so tiny ($1.3Hz$) and thus allows precision measurement in recent years\cite{9}:
\begin{equation}\label{32}
\Delta E_{H}^{exp}(1S-2S)=2466061413187.34(84)\;kHz
\end{equation}
The isotope shift of $1S-2S$ transition between $H$ and $D$ had been measured first by Schmidt-Kalar \etal \cite{10} and
quoted in \cite{11} as:
\begin{equation}\label{33}
\Delta E_{D-H}^{exp}(2S-1S)=670994337(22)\;kHz
\end{equation}
Theoretically, the above accurate data cannot be explained by the original Dirac equation with nucleus having mass $m_N\to\infty$. We propose a reduced Dirac equation ($RDE$) with electron mass replaced  by reduced mass for  $H$ and $D$ being respectively
\begin{equation}\label{34}
\mu_H=\dfrac{m_em_p}{m_e+m_p},\quad \mu_D=\dfrac{m_em_d}{m_e+m_d}
\end{equation}
Then theoretically, the $RDE$ predicts:
\begin{equation}\label{35}
\Delta E_H^{RDE}(2S-1S)=2.466067984\times 10^{15}\;Hz
\end{equation}
\begin{equation}\label{36}
\Delta E_{D-H}^{RDE}(2S-1S)=6.7101527879\times 10^{11}\;Hz
\end{equation}
which are larger than the experimental values by only $3\times10^{-6}$ and $3\times10^{-5}$ respectively. Further
radiative corrections on Eq.(\ref{35}) will be sensitive to the choice of schemes in Table $I$, the best one is
$\Delta E_H^{Theor<SV>}(2S-1S)$, deviating from the experimental data, Eq.(\ref{32}), by $-1\times10^{-7}$ only. On the other hand, besides Eq.(\ref{36}), the $\Delta E_{D-H}^{Theor}(2S-1S)$ is influenced considerably by the nuclear size effect and so less sensitive to the scheme choice of the smaller radiative correction, bringing the discrepancy between theory
and experimental data, Eq.(\ref{33}), from  $3\times10^{-5}$ down to $3\times10^{-6}$ approximately.

\section{Renormalization Group Equation ($RGE$) for $QED$ and Its Solution}
\label{sec:renormalization}

\vskip 0.1cm

In $QED$, the $FDI$ for photon self-energy (\ie, vacuum polarization) can also be evaluated \cite{4}, bringing the
charge $e$ into its renormalized one:
\begin{equation}\label{37}
e^2\to e^2_R=Z_3e^2
\end{equation}
\begin{equation}\label{38}
Z_3=1+\dfrac{\alpha}{3\pi}\left(\ln\dfrac{m^2}{\mu^2_3}-\dfrac{q^2}{5m^2}+\cdots\right)
\end{equation}
Here $m$ is the fermion (say, electron) mass, $q$ is the momentum of photon and $\mu_3$ is an arbitrary constant emerging from the treatment on the divergence like that in Eqs.(\ref{9})-({12}). Although Eq.(\ref{37}) looks like that in previous theories, it is really a new one: $e$ is the observed (physical) charge, not a "bare charge", and $Z_3$ remains finite.

The vertex function between two fermions' momenta $p$ and $p'$ with $p'-p=q$ will give another $Z_1$ \cite{4,7}. Adding
all the $FDIs$, we find the renormalized charge being:
\begin{equation}\label{39}
e_R=\dfrac{Z_2}{Z_1}Z_3^{1/2}e
\end{equation}
But the Ward-Takahashi identity ($WTI$) implies that\cite{6}
\begin{equation}\label{40}
Z_1=Z_2
\end{equation}
Hence Eq.(\ref{37}) remains valid and
\begin{equation}\label{41}
e_R(Q)=e\left\{1+\dfrac{\alpha}{2\pi}\left[\dfrac{1}{3}\ln\dfrac{m^2}{\mu^2_3}+\dfrac{1}{15}\dfrac{Q^2}{m^2}+\cdots\right]\right\}
\end{equation}
where $Q^2=-q^2>0$, with $Q$ being the 3-dimensional momentum transfer at fermion collision. The observed charge should be defined at $Q\to 0$ (Thomson scattering limit):
\begin{equation}\label{42}
e_{obs}=e_R|_{Q=0}=e
\end{equation}
which dictates that
\begin{equation}\label{43}
\mu_3=m
\end{equation}
As usual, the beta function is defined as
\begin{equation}\label{44}
\beta(\alpha,Q)\equiv Q\dfrac{\partial}{\partial Q}\alpha_R(Q)
\end{equation}
From Eq.(\ref{41}), it is found in \cite{4} that
\begin{equation}\label{45}
\beta(\alpha,Q)=\dfrac{2\alpha^2}{3\pi}-\dfrac{4\alpha^2m^2}{\pi Q^2}\left[1+\dfrac{2m^2}{\sqrt{Q^4+4m^2Q^2}}
\ln\dfrac{\sqrt{Q^4+4m^2Q^2}-Q^2}{\sqrt{Q^4+4m^2Q^2}+Q^2}\right]
\end{equation}
\begin{equation}\label{46}
\beta(\alpha,Q)=\dfrac{2\alpha^2}{15\pi}\dfrac{Q^2}{m^2},\quad (\dfrac{Q^2}{m^2}\ll1)
\end{equation}
Evidently, the $e_R(Q)$ will increase with $Q$, becoming a running coupling constant ($RCC$). To calculate it, usually
a renormalization-group-equation ($RGE$) was derived for $QED$ by setting $\alpha\to\alpha_R(Q)$ and $Q\to\infty$ in
$\beta(\alpha,Q)$ yielding:
\begin{equation}\label{47}
Q\dfrac{d}{dQ}\alpha_R=\dfrac{2\alpha_R^2}{3\pi}
\end{equation}
with its solution
\begin{equation}\label{48}
\alpha_R(Q)=\dfrac{\alpha}{1-\frac{2\alpha}{3\pi}\ln\frac{Q}{m}}
\end{equation}
Here, the renormalization was forced to be made at $Q^2=m^2$ so that
\begin{equation}\label{49}
\alpha_R|_{Q^2=m^2}=\alpha
\end{equation}
This is inconsistent with the physical condition, Eq.(\ref{42}), a defect due to ignoring the mass $m$, which plays a
dominant role at low $Q$ region as shown by the Eq.(\ref{46}). As an improvement, a more practical $RGE$ is constructed
in \cite{4} by changing $\alpha$ into $\alpha_R(Q)$ in Eq.(\ref{45}) for the entire $Q$ region and adding up contributions from $9$ elementary charged
fermions ($e,\mu,\tau,u,d,s,c,b,t$). Then $\alpha_R$ can be numerically calculated as a function of $\ln(Q/m_e)$,
starting from
\begin{equation}\label{50}
\alpha_R(Q=0)=\alpha=(137.03599)^{-1}
\end{equation}
and passing through another experimental data point \cite{12}
\begin{equation}\label{51}
\alpha_R(Q=M_Z=91.1880\,GeV)=(128.89)^{-1}
\end{equation}
In this way, after adopting three heavy quarks' masses as $m_c=1.031\,GeV,\; m_b=4.326\,GeV$ (see section 9.5D in \cite{13}) and $m_t=175\,GeV$, three light quarks masses
\begin{equation}\label{52}
m_u=8\,MeV,\;m_d=10\,MeV,\;m_s=200\,MeV
\end{equation}
(or averaged mass for $u,d,s$ being $92\ MeV$) can be fitted as shown in Fig.1 of Ref.\cite{4}.

\section{$RGE$ for $QCD$, Threshold Energies of Quarks Hadronization}
\label{sec:hadronization}

\vskip 0.1cm

Different from the $RCC$ in $QED$, the $RCC$ in $QCD$, denoted by $\alpha_s(Q)=\frac{1}{4\pi}g_s^2(Q)$, is much larger and
decreases with the increase of $Q$. Usually, the $RGE$ for $QCD$ reads (see p.551-552 in \cite{2})
\begin{equation}\label{53}
Q\dfrac{d}{dQ}\alpha_s(Q)=\beta(\alpha_s(Q))=-\dfrac{\beta_0}{2\pi}\alpha_s^2(Q)
\end{equation}
where
\begin{equation}\label{54}
\beta_0=11-\dfrac{2}{3}n_f
\end{equation}
with $n_f$ being the number of quarks' flavor. Eq.(\ref{53}) looks similar to Eq.(\ref{47}), but the negative sign in beta
function implies the property of asymptotic freedom in the strong interaction. Solution of Eq.(\ref{53}) is give by
Eq.(\ref{6}), also bears some resemblance to Eq.(\ref{48}).

Let us make a comparison between Eqs.(\ref{6}) and (\ref{48}). Besides the difference in sign ("$+$" versus "$-$") in the
denominator, there is another one: The mass scale $\mu$ remains arbitrary in $QCD$ whereas $m$ in $QED$ means the mass of an
observed charged fermion (usually electron).

As we already see, even for $QED$, one mass ($m_e$) is far from enough, let alone in $QCD$, where quarks' masses are much
heavier. How can we ignore them?

Usually, to remove the arbitrary mass scale $\mu$, a parameter $\Lambda_{QCD}$ is often defined via equation
\begin{equation}\label{55}
\dfrac{\beta_0}{2\pi}\alpha_s(\mu)\ln\left(\dfrac{\mu}{\Lambda_{QCD}}\right)=1
\end{equation}
such that a simpler formula for $\alpha_s(Q)$ can be found as
\begin{equation}\label{56}
\alpha_s(Q)=\dfrac{2\pi}{\beta_0\ln(Q/\Lambda_{QCD})}
\end{equation}
Then the precision experimental data \cite{1}
\begin{equation}\label{57}
\alpha_s(M_Z=91.1876\,GeV)=0.1176
\end{equation}
serves as a substitute for $\alpha_s(\mu)$ in Eq.(\ref{55}), yielding
\begin{equation}\label{58}
\Lambda_{QCD}=M_Z\exp\left[\dfrac{-2\pi}{\alpha_s(M_Z)\beta_0}\right]
=\left\{\begin{array}{l}
          240\;MeV,\;(n_f=3) \\
          150\;MeV,\;(n_f=4) \\
          85.8\;MeV,\;(n_f=5) \\
          44.2\;MeV,\;(n_f=6)
        \end{array}\right.
\end{equation}
It is evident from Eq.(\ref{56}) that
\begin{equation}\label{59}
\alpha_s(\Lambda_{QCD})\to\infty
\end{equation}
which implies the "infrared confinement" of quarks.

However, the value of $\Lambda_{QCD}$ sensitively depends on the flavor number $n_f$ as shown by Eq.(\ref{58}) but
is independent of the concrete flavor of a quark under consideration. Moreover, the divergence of $\alpha_s$ appears
at $\Lambda_{QCD}$. These features seem not so reasonable and are not consistent with the experimental fact that the lighter
a quark's mass is, the lower its "threshold energy for hadronization" will be.

The way out of above difficulties is clear. Just like in Eq.(\ref{45}) for $QED$, where rather than just the first term,
all terms for $9$ fermions should be added, now for $RGE$ in $QCD$, all masses of $6$ quarks should be preserved. In this
way, the $\alpha_{si}(Q)$ are numerically calculated for $i=u,d,s,c,b$ respectively in Ref.\cite{4}. Starting from
$\alpha_s(M_Z)=0.118$ (the common renormalization point), their running curves (as shown by Figs.2 and 3 in \cite{4}) follow the trend of experimental data (as shown on p.158 of Ref.\cite{1}) quite well but separate at the low $Q$ region. Each of
them rises to a maximum $\alpha_{si}^{max}$ at $Q=\Lambda_i$ and then suddenly drops to zero at $Q=0$.

For example, for $b$ quark, $\Lambda_b=7.04\, GeV,\;\alpha_{sb}^{max}=0.161$. If tentatively explain $L_b\sim \hbar/\Lambda_b$
as some critical length scale of $b\bar{b}$ pair, then $L_b\sim 0.02805\,fm$. In \cite{4}, it is further guessed that
$E_b^{theor}\sim \alpha_{sb}^{max}/L_b\sim 1.133\, GeV$ being the order of excitation energy for breaking the binding
$b\bar{b}$ pair, \ie, the hadronization threshold energy of Upsilon $\Upsilon(b\bar{b})$ against its dissociation into
two bosons. It is indeed the case found experimentally\cite{1}:
\begin{equation}\label{60}
M(\Upsilon(4S))-M(\Upsilon)=1.12\,GeV,\quad M(\Upsilon(4S))\to B^+B \;\text{or}\;B^0\bar{B^0}
\end{equation}
Similarly, we can estimate from Ref.\cite{4} that $E_c^{theor}\sim0.398\,GeV$ which also corresponds to experiments that\cite{1}
\begin{equation}\label{61}
M(\psi(3770))-M(\psi(3097))=673\,MeV,\quad \psi(3770)\to D^+D^-\;\text{or}\;D^0\bar{D^0}
\end{equation}
It seems that $E_s^{theor}\sim90\,MeV$ and $E_{u,d}^{theor}\sim 0.4\,MeV$ are not so reliable but still reasonable.

\section{$\lambda \phi^4$ Model and Higgs Mass in the Standard Model}
\label{sec:higgsmass}

\vskip 0.1cm

The Lagrangian density of $\lambda\phi^4$ ($\phi(x)$ is a real scalar field) model is defined by
\begin{equation}\label{62}
{\cal L}=\dfrac{1}{2}\partial_\mu\phi\partial^\mu\phi+\dfrac{1}{2}\sigma\phi^2-\dfrac{\lambda}{4!}\phi^4
\end{equation}
The importance of this model lies in the "wrong sign" of mass term ($\sigma=-m^2>0$) which leads to the spontaneous
symmetry breaking ($SSB$) at the tree level ($L=0$). The effective potential ($EP$) reads
\begin{equation}\label{63}
V_0(\phi)=-\dfrac{1}{2}\sigma\phi^2+\dfrac{\lambda}{4!}\phi^4
\end{equation}
(The subscript "$0$" refers to $L=0$). Obviously, $V_0(\phi)$ has two extremum, one is a maximum:
\begin{equation}\label{64}
\phi_0=0\quad\text{(symmetric phase)}
\end{equation}
while the other one is a minimum:
\begin{equation}\label{65}
\phi_1^2=\dfrac{6\sigma}{\lambda}\quad\text{(SSB phase)}
\end{equation}
At the $QFT$ level, the $EP$ evolves into
\begin{equation}\label{66}
V=V_0+V_1+\cdots
\end{equation}
The theory for $EP$ had been developed by various authors \cite{14,15,16,17,18}, with $L=1$ contribution to $EP$ being
evaluated as:
\begin{equation}\label{67}
V_1(\phi)=\dfrac{1}{2}\int\dfrac{d^4k_E}{(2\pi)^4}\ln(k_E^2-\sigma+\frac{1}{2}\lambda\phi^2)
\end{equation}
This highly divergent integral (in 4-dimensional Euclidean momentum space) is treated in Ref.\cite{19} like that for
Eqs.(\ref{10})-({12}). First, three times of differentiation with respect to $M^2=-\sigma+\frac{1}{2}\lambda\phi^2$
are needed before it becomes just convergent.
\begin{equation}\label{68}
\dfrac{\partial^3V_1}{\partial(M^2)^3}=\int\dfrac{d^4k_E}{(2\pi)^4}\dfrac{1}{(k_E^2+M^2)^3}=\dfrac{1}{2(4\pi)^2M^2}
\end{equation}
Second, three times integration with respect to $M^2$ are performed, yielding
\begin{equation}\label{69}
V_1(\phi)=\dfrac{1}{2(4\pi)^2}\left\{\dfrac{M^4}{2}(\ln M^2-\dfrac{1}{2})-\dfrac{1}{2}M^4+\dfrac{1}{2}C_1M^4+C_2M^2+C_3\right\}
\end{equation}
As expected, three arbitrary constants $C_1,C_2,C_3$ appear. The renormalization amounts to fix them at our disposal.

Third, like that in Eq.(\ref{3}), for eliminating the ambiguity of dimension in the first term involving $\ln M^2$,
the only possible choice of $C_1=-\ln\mu^2$ is fixed. Then the choice of $C_2=\mu^2=2\sigma$ and $C_3=-\sigma^2+(4\pi)^2\frac{3\sigma^2}{\lambda}$ leads to $V=V_0+V_1$ with its derivatives being given at the Table II
\cite{19}.

\renewcommand\arraystretch{1.5}
\begin{center}
\hspace*{-16mm}\begin{tabular}{|c|c|c|}
\multicolumn{2}{c}{Table II. Effective potential of $\lambda\phi^4$ model with $SSB$}\tabularnewline
  \hline
    & $SSB$ phase & symmetric phase \tabularnewline
  \hline
  $\phi$ & $\phi_1=\sqrt{\dfrac{6\sigma}{\lambda}}$  & $\phi_0=0$ \tabularnewline
  \hline
  $V$ & $0$  & $-\dfrac{\sigma^2}{2(4\pi)^2}\left[\dfrac{15}{4}+\dfrac{1}{2}\ln2-i\dfrac{\pi}{2}\right]+\dfrac{3}{2}\dfrac{\sigma^2}{\lambda}$ \tabularnewline
   \hline
  $\dfrac{dV}{d\phi}$ & $0$ & $0$ \tabularnewline
   \hline
  $\dfrac{d^2V}{d\phi^2}$ & $2\sigma$  & $-\sigma\left[1-\dfrac{\lambda}{2(4\pi)^2}(3+\ln2-i\pi)\right]$\tabularnewline
   \hline
  $\dfrac{d^3V}{d\phi^3}$ & $\lambda\sqrt{\dfrac{6\sigma}{\lambda}}\left[1+\dfrac{3\lambda}{2(4\pi)^2}\right]$  & $0$ \tabularnewline
   \hline
 $\dfrac{d^4V}{d\phi^4}$ & $\lambda\left[1+\dfrac{9\lambda}{2(4\pi)^2}\right]$  & $\lambda\left[1-\dfrac{3\lambda}{2(4\pi)^2}(\ln2-i\pi)\right]$ \tabularnewline
  \hline
\end{tabular}
\end{center}

Note that, with the above assignment of $C_i\,(i=1,2,3)$, both the position of the $SSB$ phase, $\phi_1$, and the mass
$m_\sigma$ excited above it take the same expression as that at the tree level
\begin{equation}\label{70}
m_\sigma^2=\dfrac{d^2V}{d\phi^2}|_{\phi=\phi_1}=2\sigma
\end{equation}
However, the renormalization coupling constant
\begin{equation}\label{71}
\lambda_R=\dfrac{d^4V}{d\phi^4}|_{\phi=\phi_1}=\lambda\left[1+\dfrac{9\lambda}{2(4\pi)^2}\right]
\end{equation}
does receive some quantum correction on its classical value $\lambda$. Hence it is suddenly realized that the invariant meaning of $\lambda$ in the Lagrangian, Eq.(\ref{62}), is by no means a "coupling constant", but the ratio of two mass scales \cite{21}.
\begin{equation}\label{72}
\lambda=3\dfrac{m_\sigma^2}{\phi_1^2}
\end{equation}
Two parameters, $\sigma$ and $\lambda$, together with Eqs.(\ref{70}) and (\ref{72}), should all be preserved through out
high loop ($L$) evaluations in perturbation theory until $L\to\infty$, \ie, in any nonperturbative treatment.

On the other hand, the above assignment of $C_i$ renders the appearance of imaginary part in $V$ and its derivatives at
the symmetric phase ($\phi_0=0$). It means the instability of symmetric phase at the presence of stable $SSB$ phase.

It is interesting to see that an alternative choice of
\begin{equation}\label{73}
C_1=-\ln(-\sigma),\;C_2=-\sigma,\;C_3=-\dfrac{1}{4}\sigma^2
\end{equation}
would leads to the survival of $\phi_0=0$ as a semistable state with
\begin{equation}\label{74}
V(0)=\dfrac{dV}{d\phi}|_{\phi=0}=\dfrac{d^3V}{d\phi^3}|_{\phi=0}=0
\end{equation}
\begin{equation}\label{75}
\dfrac{d^2V}{d\phi^2}|_{\phi=0}=-\sigma,\;\dfrac{d^4V}{d\phi^4}|_{\phi=0}=\lambda
\end{equation}
whereas no real $SSB$ solution exists. Hence we see that two different choices of $C_i$ lead two separable sectors in
the effective potential \cite{19}.

In 1989, we had estimated the upper and lower bounds of Higgs mass $M_H$ in the standard model of particle physics by
using a nonperturbative approach in $QFT$ --- the Gaussian effective potential ($GEP$) method, yielding\cite{20}
\begin{equation}\label{76}
76\,GeV<M_H<170\,GeV
\end{equation}
Like many authors, we were bothered a lot by divergences. After a deeper understanding on $\lambda\phi^4$ model\cite{19},
this problem was restudied in 1998 by a combination of $GEP$ with our $RRM$, yielding
\begin{equation}\label{77}
M_H=138\,GeV
\end{equation}
which is based on the following input of experimental data:
\begin{equation}\label{78}
\begin{array}{l}
  M_W=80.359\,GeV,\;M_Z=91.1884\,GeV, \\
  \alpha^{-1}=\dfrac{4\pi}{g^2\sin^2\theta_W}=128.89,\;\sin^2\theta_W=0.2317
\end{array}
\end{equation}
where $\theta_W$ is the weak mixing (Weinberg) angle. As now the search for Higgs particle becomes so urgent experimentally
while the theoretical estimation about its mass still remains uncertain\cite{1}, we think our method \cite{21} with its
prediction, Eq.(\ref{77}), deserves to be reconsidered.

\section{Summary and Discussion}
\label{sec:discussion}

\vskip 0.1cm

Our $RRM$ was first proposed by J. F. Yang in 1994\cite{22}, then elaborated in a series of papers since 1998 (\cite{19,21,
23,24,4,7} \etc). What we have been thinking about is: Why the divergence emerges inevitably in $QFT$? What is the essential meaning of a regularization and renormalization procedure?

For instance, in a pioneering work to explain the Lamb shift, Welton (\cite{25}, see also section 9.6B in Ref.\cite{13})
encountered a divergent integral $I=\int\frac{d\omega}{\omega}$ with $\omega$ being the (angular) frequency of virtual photon (vacuum fluctuation). He simply set the lower and upper bounds by $\omega_{min}\sim mZ\alpha=\frac{Z}{a}$ ($a$ is Bohr
radius) and $\omega_{max}\sim m$ respectively, arriving at $I\simeq \ln(\frac{1}{Z\alpha})=4.92\,(Z=1)$ which leads to
an estimation of Lamb shift $L_H^{theor}(2S_{1/2}-2P_{1/2})\simeq 668\,MHz$. However, if instead of Bohr radius, the lower cutoff is provided by the electron's binding energy, one would get $I\simeq \ln(Z\alpha)^{-2}$ and
$L_H^{theor}(2S_{1/2}-2P_{1/2})\simeq 1336\,MHz$ (see Eq.(\ref{30}) in the Ref.\cite{26}). We see the integral $I$ being
a dimensionless number, not very large ($I\leq 10$), but uncertain indeed. The root cause of uncertainty lies in the fact
that a reconfirmation process of electron mass like Eq.(\ref{20}) was missing.

For further clarity, a study on Lamb shift in the form of noncovariant $QED$ was performed in \cite{24} (see also Appendix 9A in \cite{13}). Beginning with Eqs.(\ref{18}) and (\ref{19}) (with $m\to\mu=\frac{mm_N}{m+m_N}$), the perturbative
calculation of electron's self-energy at second order in noncovariant form (corresponding to the one-loop ($L=1$) order in
covariant form) leads to an energy increase of an electron with momentum $\bf p$, $\Delta E_p$, which contains divergence
and can be handled just like that in Eqs.(\ref{10}) and (\ref{67}), yielding
\begin{equation}\label{79}
\begin{array}{l}
  \Delta E_p=b_1p^2+b_2p^4+\cdots\\[5mm]
  b_1=\dfrac{\alpha}{\pi\mu}(\dfrac{4}{3}\ln2+\dfrac{4}{3}\ln\mu-\dfrac{4}{3}C_1)\\[5mm]
  b_2=\dfrac{\alpha}{\pi\mu^3}(-\dfrac{2}{15})
\end{array}
\end{equation}
The only choice of arbitrary constant $C_1$ is to make $b_1=0$ such that the reduced mass $\mu$ in Eq.(\ref{18}) can be
reconfirmed. However, Eq.(\ref{19}) must be supplemented by the interaction between electron's spin and the radiation field
\begin{equation}\label{80}
H'_{int}=\dfrac{ge\hbar}{4\mu c}{\boldsymbol \sigma}\cdot\nabla\times{\bf A}
\end{equation}
where $g=2\times 1.0011596522$ is gyromagnetic ratio of electron. Similar treatment leads to
\begin{equation}\label{81}
\begin{array}{l}
  \Delta E'_p=b'_0+b'_1p^2+b'_2p^4+\cdots\\[4mm]
  b'_0=\dfrac{g^2}{4}\dfrac{\alpha\mu}{\pi}[4(\ln2+\ln\mu)-4C_2-\dfrac{2C_3}{\mu}-\dfrac{C_4}{\mu^2}]\\[4mm]
  b'_1=\dfrac{g^2}{4}\dfrac{\alpha}{\pi\mu}(\dfrac{4}{3}\ln2+2+\dfrac{4}{3}\ln\mu-\dfrac{4}{3}C_2)\\[4mm]
  b'_2=\dfrac{g^2}{4}\dfrac{\alpha}{\pi\mu^3}(-\dfrac{1}{15})
\end{array}
\end{equation}
Because $\mu$ has already been reconfirmed (by $b_1=0$), the only choice of arbitrary constant $C_2$ is to cancel the
ambiguous term with $\ln\mu,\;C_2=\ln\mu$, leaving a nonzero $b'_1p^2$ and combining with $\frac{1}{2\mu}{\bf p}^2$. Hence
$\mu$ really acquires a modification as
\begin{equation}\label{82}
\mu\to \mu_{obs}=\dfrac{\mu}{1+\beta},\qquad \beta=\dfrac{g^2\alpha}{2\pi}(\dfrac{4}{3}\ln2+2)
\end{equation}
where $\mu_{obs}=\frac{m_em_N}{m_e+m_N}$. Then constants $C_3$ and $C_4$ must be chosen such that $b'_0=0$.

Notice that, however, the spin induced interaction, Eq.(\ref{80}), endows electron with relativistic feature, creating
a term ($-\frac{1}{8\mu^3}p^4$) in its kinetic energy. Yet the modification on $\mu$ shown in Eq.(\ref{82}) does
induce a corresponding change $-\frac{1}{8}(\frac{1}{\mu_{obs}^3}-\frac{1}{\mu^3})p^4$, which should be regarded as an invisible "background" and subtracted from the $p^4$ term induced by the radiative corrections. Hence the "renormalized"
$b_2$ should be
\begin{equation}\label{83}
b_2^R=b_2+b'_2+\frac{1}{8\mu^3}(3\beta+3\beta^2+\beta^3)\simeq \dfrac{\alpha}{\pi\mu_{obs}^3}(1.99808)
\end{equation}
where only the lowest approximation is kept ($\mu\simeq\mu_{obs}$). Hence the radiative correction on the energy
level of a stationary state $|Z,n,l\rangle$ in hydrogenlike atom simply reads
\begin{equation}\label{84}
\Delta E^{rad}(Z.n,l)=\langle Z,n,l|b_2^Rp^4|Z,n,l\rangle=[\dfrac{8n}{3l+1}-3]\dfrac{b_2^RZ^4\alpha^4}{n^4}\mu_{obs}^4
\end{equation}
This contribution, together with that from the vacuum polarization (borrowed from covariant theory) and nuclear size effect, gives a theoretical value for the Lamb shift \cite{24}
\begin{equation}\label{85}
L_H^{theor}(2S_{1/2}-2P_{1/2})= 1056.52\,MHz
\end{equation}
which is smaller than the experimental value, Eq.(\ref{29}), by $0.13\%$ only.

Despite the approximation involved, the above method clearly shows that our regularization is by noo means a trick to curb the divergence. Rather, it is a natural way to transform a divergence into some arbitrary constants, revealing that the essential meaning of divergence is just the "uncertainty" in the theory. Thus so-called renormalization turn out to be nothing but a process of reconfirmation to fix these constants via experiments. We must reconfirm a mass before it could
be modified via radiative corrections. Either "skipping over the first step" or "combining two steps into one" is not allowed.

In deeper understanding, our $RRM$ is based on a "principle of relativity" in epistemology\cite{27}: Every thing is
moving and becomes recognizable only in relationship with other things. What we can understand is either no mass scale
or two mass scales, but never one mass scale. This scenario is clearly displayed in the Gross-Neveu model\cite{28}, also
in the $\lambda\phi^4$ model with $SSB$ as shown by Eqs.(\ref{65}), (\ref{70}) and (\ref{72}). The vacuum expectation
value of field, $\phi_1$, just provides a "mass unit" for the mass $m_\sigma$ excited on the vacuum with $SSB$.

Similarly, in perturbative $QFT$, we will be able to calculate various radiative corrections on a particle only when its mass $m$ can be reconfirmed again and again throughout any high loop ($L$) order of theory until $L\to\infty$. Just
like one has to reconfirm his plane ticket before his departure from the airport, he  must use the same name throughout
his entire jouney\cite{4}.

\section*{Acknowledgements}

\vskip 0.1cm

We thank S. Q. Chen, Y. S. Duan, R. T. Fu, S. S. Feng, T. Huang, P. T. Leung, W. F. Lu,
X. T. Song, F. Wang, H. B. Wang, K. Wu, Y. L. Wu, J. Yan, G. H. Yang,
J. F. Yang and Z. X. Zhang for close collaborations and/or helpful discussions.




\begin{thebibliography}{99}

\bibitem{1}
C. Amsler \etal,, \plb{667}{2008}{1} We refer to the Particle Physics Booklet extractd from it.

\bibitem[2]{2}
M. E. Peskin and D. V. Schroeder, {\it An Introduction to Quantum Field
Theory}, (Addison-Wesley Publishing Company, 1995).

\bibitem[3]{3}
J. C. Collins, {\it Renormalization}, (Cambridge University Press, Cambridge, 1984)

\bibitem[4]{4}
G. J. Ni, G. H. Yang and R. T.Fu, Int. J. Mod. Phys. A{\bf 16}, 2873(2001)

\bibitem[5]{5}
J. J. Sakurai, {\it Advanced Quantum Mechanics} (Addison-Wesley Publishing Company, 1967)

\bibitem[6]{6}
C. Itzykson and J-B. Zuber, {\it Quantum Field Theory} (McGraw-Hill Book Company, 1980)

\bibitem[7]{7}
G. J. Ni, J. J. Xu and S. Y. Lou, Submitted to Chinese Physics B, {\tt quant-ph}/0511197.

\bibitem[8]{8}
H. A. Bethe, \pr{72}{1947}{339}.

\bibitem[9]{9}
Th. Udem \etal, \prl{79}{1997}{2646}.

\bibitem{10}
F. Schmidt-Kaler \etal, \prl{70}{1993}{2261}

\bibitem[11]{11}
M. Weitz \etal, \pra{52}{1995}{2664}.

\bibitem[12]{12}
H. Burkhardt and B. Pietrzyk, \plb{356}{1995}{398}

\bibitem[13]{13}
G. J. Ni and S. Q. Chen, {\it Advanced Quantum Mechanics}, 2nd
Edition (Fudan University Press, 2003); English Edition was
published by Rinton Press, 2002.

\bibitem[14]{14}
S. Coleman and E. Weinberg, \prd{7}{1973}{1883}

\bibitem[15]{15}
R. Jackiw, \prd{9}{1974}{1686}

\bibitem[16]{16}
C. W. Bernard, \prd{9}{1974}{3312}

\bibitem[17]{17}
L. Dolan and R. Jackiw, \prd{9}{1974}{3320}

\bibitem[18]{18}
S. Weinberg, \prd{9}{1974}{3357}

\bibitem[19]{19}
G. J. Ni and S. Q. Chen, Acta Physica Sinica (Overseas Edition), {\bf
7}, 401 (1998).

\bibitem[20]{20}
S. Y. Lou and G. J. Ni, \prd{40}{1989}{3040}

\bibitem[21]{21}
G. J. Ni, S. Y. Lou, W. F. Lu and J. F. Yang, Science in China
(Series A), {\bf 41}, 1206 (1998), \hepph{9801264}.

\bibitem{22}
J. F. Yang, Thesis for PhD (Fudan University, 1994); hep-th/9708104;\\
J. F. Yang and G. J. Ni, Acta Physica Sinica (Overseas Edition), {\bf
4}, 88 (1995).

\bibitem[23]{23}
S. S. Feng and G. J. Ni, Int. J. Mod. Phys. A, {\bf 14}, 4259
(1999).

\bibitem[24]{24}
G. J. Ni, H. B. Wang, J. Yan and H. L. Li, High Energy Physics and
Nuclear Physics, {\bf 24}, 400 (2000).

\bibitem[25]{25}
T. A. Welton, \pr{74}{1948}{1157}

\bibitem[26]{26}
M. I. Eides, H. Grotch and V. A. Shelyuto, \prep{342}{2001}{63}

\bibitem[27]{27}
G. J. Ni, Principle of Relativity in Physics and in Epistemology, in {\it Relativity, Gravitation, Cosmology: New development}, (NOVA Science Publisher, to be published)

\bibitem[28]{28}
D. J. Gross and A. Neveu, \prd{10}{1974}{3235}

\bibitem[29]{29}
G. J. Ni, S. Q. Chen, J. J. Xu and S. Y. Lou, {\it Essence of special
relativity, reduced Dirac equation and antigravity}, Preprint.


\end{thebibliography}
\end{document}